\documentclass[12pt]{iopart}

\usepackage{graphicx}

\begin{document}

\title{A demonstration device for cosmic rays telescopes}

\author{Salvatore Esposito}

\address{Istituto Nazionale di Fisica Nucleare, Sezione di Napoli, Complesso Universitario di Monte
S.\,Angelo, via Cinthia, I-80126 Naples, Italy}
\ead{Salvatore.Esposito@na.infn.it}
\vspace{10pt}

\begin{abstract}
We describe a hands-on accurate demonstrator for cosmic rays realized by six high school students, whose main aim is to show the relevance and the functioning of the principal parts of a cosmic rays telescope (muon detector), with the help of two large size wooden artifacts. The first one points out how cosmic rays can be tracked in a muon telescope, while the other one shows the key avalanche process of electronic ionization that effectively allows muon detection through a photomultiplier. Incoming cosmic rays are visualized in terms of laser beams, whose 3D trajectory is highlighted by the turning on of LEDs on two orthogonal matrices. Instead the avalanche ionization process is demonstrated through the avalanche falling of glass marbles on an inclined plane, finally turning on a LED. A pictured poster accompanying the demonstrator is as well effective in assisting cosmic rays demonstration and its detection. The success of the demonstrator has been fully proven by general public during a Science Festival, the corresponding project winning the Honorable Mention in a dedicated competition.
\end{abstract}

%
%
%
%
%

\section{Introduction}

\begin{figure}[t]
\begin{center}
\begin{tabular}{c}
\includegraphics[width=10cm]{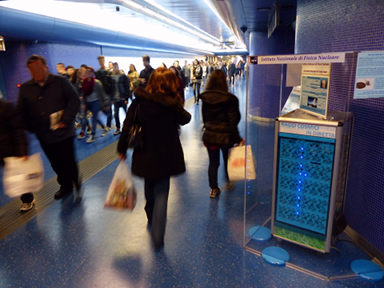} 
\end{tabular}
\caption{The I.N.F.N. installation of a cosmic rays telescope in the Toledo Metro Station in Naples.}
\label{fig11}
\end{center}
\end{figure}

\noindent Inside the ``Toledo" Metro Station in Naples (Italy), since May 2014 a scientific installation  developed by the Gran Sasso National Laboratory of the Italian National Institute of Nuclear Physics (I.N.F.N.) is present, aimed at detecting the underground cosmic radiation \cite{Pazos} at about 40 meters of depth (see Fig. \ref{fig11}). Operating as an effective cosmic muons telescope, such a compact particle tracking system \cite{Arneodo} was originally designed for didactic and outreach activities, and, along with the associated multimedia Totem providing videos on cosmic rays physics, it currently works for communication and dissemination of scientific culture in Naples and its surroundings.

The cosmic rays telescope consists of plastic scintillator bars, which are optically coupled -- through wavelength shifter fibers embedded into each bar -- to Silicon Photomultipliers (SiPM) connected to a PCB board to be biased and read, then monitoring the working parameters and remotely connecting the detector. The whole system \cite{Pazos}, comprised of 200 electronic channels organized into 10 couples of orthogonal planes, allows the 3D reconstruction of the muons crossing the detector, and a system of two matrices of LEDs -- one for every scintillator bar triggered by charged particle interactions -- has been implemented in order to make easier the display of muon tracks to non experts. 

Such device with LED matrices has indeed had a major impact on science popularization in Naples and around, introducing people with no background in physics to the world of particle physics, just allowing them to visualize high energy particles crossing a detector. The success of this installation has been so effective that a competition was launched in 2016 by I.N.F.N. at high secondary schools, aimed at engaging teachers and students in astroparticle physics projects, directly related or not to cosmic rays. A number of students participated to such competition, preparing posters, multimedia presentations, artifacts and even didactic experiments, a selection of which has been presented to public during the renowned science festival  ``Futuro Remoto" in May 2017, with a considerable success \cite{AramoFB}. The winners of the competition have had the opportunity to visit the Gran Sasso National Laboratory.

In this paper, we will describe just one of these projects (winning the Honorable Mention of the Board of Examiners of the competition), whose core -- contrary to the current main stream -- had not a computer-based approach or multimedia presentation, but rather a physical demonstrator built with wood, glass balls, LEDs and other poor materials. Such demonstrator, along with students explanations, revealed its success in explaining to general public (from different countries, including Italy, England, Germany and even India) what are cosmic rays and how they can be detected.

\section{Cosmic rays and their detection}

\noindent The demonstrator has been realized by six students of the {\it Liceo  ``Virgilio"} in Pozzuoli (Naples),  and was aimed mainly to understand how a cosmic rays detector works, by describing its principal parts and allowing people to realize its physical operating principles. However, a preliminary work has been devoted to gather information about cosmic rays and their detection \cite{Gaisser}. Such information (collected on the web, as well as in lectures delivered by experts of I.N.F.N.) then came into a poster, just realized as a fundamental accompanying part of the demonstrator described in the following sections. 

\begin{figure}[t]
\begin{center}
\begin{tabular}{cc}
\includegraphics[width=5cm]{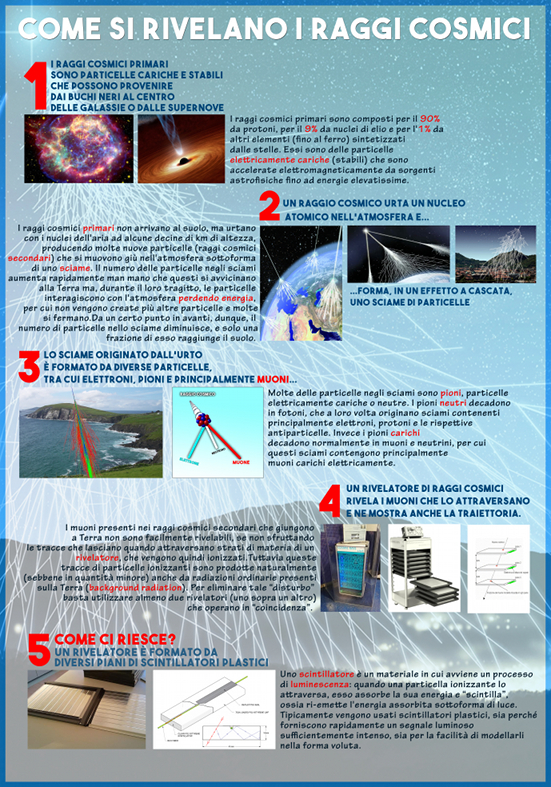} & \includegraphics[width=5cm]{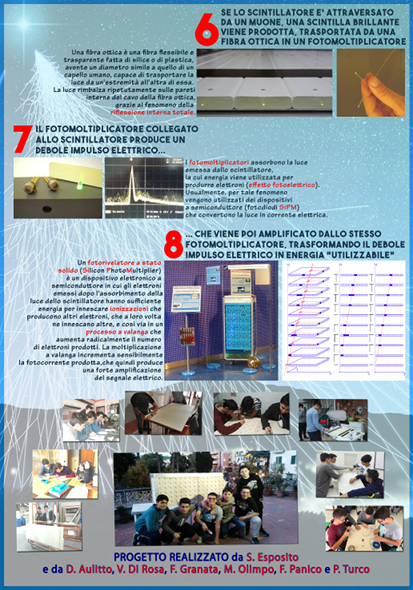} 
\end{tabular}
\caption{The original poster (in Italian) on ``How detecting Cosmic Rays" supporting the demonstrator.}
\label{fig21}
\end{center}
\end{figure}

The poster (Fig. \ref{fig21}) was extensively used by the students when explaining cosmic rays physics to the public, and served also to introduce the demonstrator.  The public was guided into the topic by means of the following steps.

\begin{itemize}
\item[{1.}]  {\bf Cosmic rays origins.}  Primary cosmic rays are stable charged particles coming from astrophysical sources such as supernovae, active galactic nuclei, and so on. They are accelerated to extremely high energies by electromagnetic fields during their journey to the Earth. They are composed by protons (90\%), alpha particles 9\%) and other elemental nuclei (till to iron) produced by stars (1\%).

\item[{2.}] {\bf Showers in the atmosphere} Primary cosmic rays reaching the Earth do not arrive to the ground, but scatter off the atomic nuclei high in the atmosphere, at several tens of kilometers, thus producing secondary cosmic rays moving  as a shower in the atmosphere. The number of particles in a shower rapidly increases as in an avalanche effect, since the energetic particles produce more particles when moving towards the ground. Their energy, however, evidently decreases during the journey in the atmosphere, so that no more particles are generated at a certain point, and some of them start to stop. From now on, the number of the particles in the shower decreases, and only a fraction of it reaches the ground.

\begin{figure}[t]
\begin{center}
\begin{tabular}{cc}
\includegraphics[width=7.5cm, height=4cm]{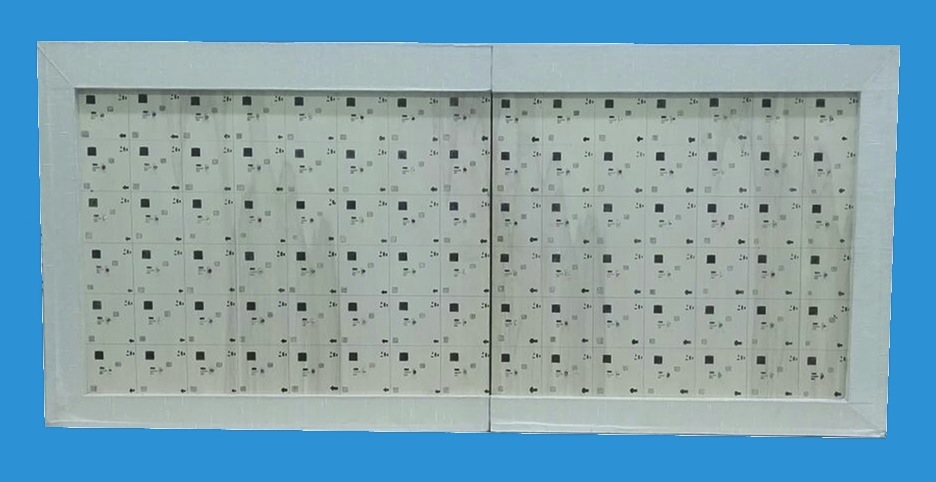} & \includegraphics[width=7.5cm, height=4cm]{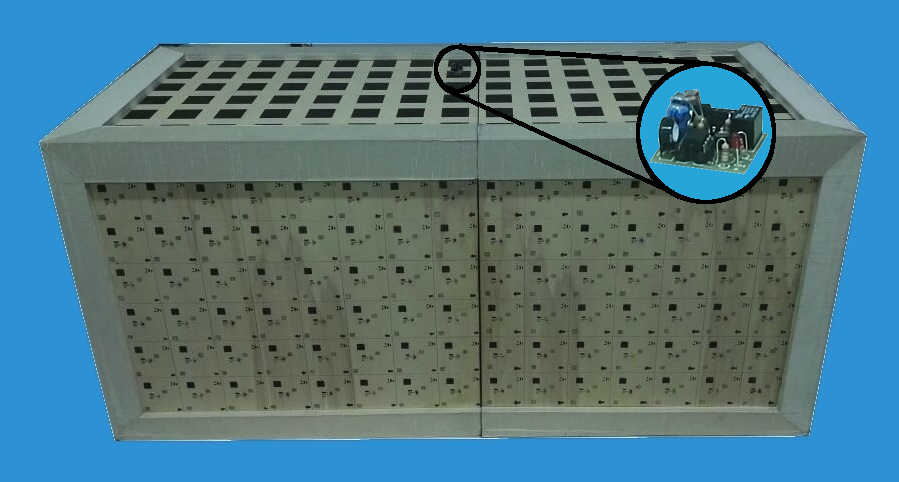} \\
\includegraphics[width=7.5cm, height=4cm]{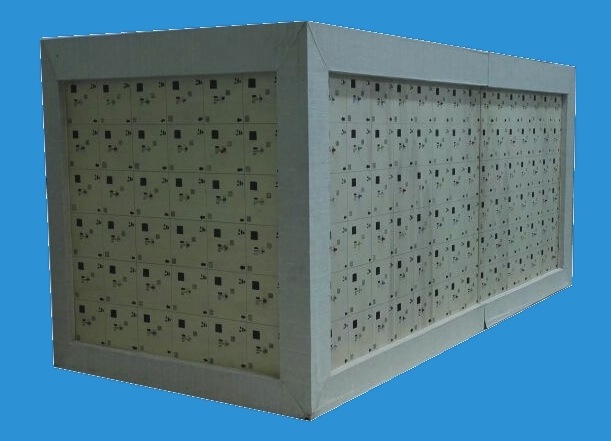} & \includegraphics[width=7.5cm, height=4cm]{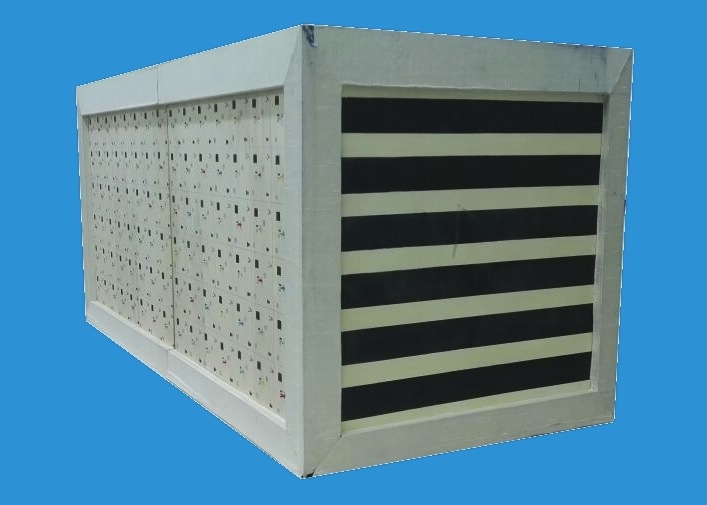}
\end{tabular}
\caption{The model of a cosmic rays telescope. On the front and on the left sides of the model, the typical elements of the front-end electronics are drawn. On the right side and on the top one, instead, white bars are sketched, denoting the scintillator bars of a cosmic rays telescope.}
\label{fig31}
\end{center}
\end{figure}

\item[{3.}] {\bf Composition of the secondaries.} Primary cosmic rays produce mainly pions when scattering off the atmosphere. Neutral pions decays into photons, that in turn gives origin to showers containing mainly electrons, protons and their corresponding antiparticles, while charged pions usually decays into muons and neutrinos. At Earth ground, the most abundant charged particles (which can be easily detected) are then muons. 

\item[{4.}] {\bf Detecting muons at the ground.} Secondary cosmic muons can be displayed by the tracks left in the different layers of matter in a detector, due to ionization processes induced by the charged muons. However, similar tracks are left also by other ionizing particles (although to a lesser extent)  naturally produced by ordinary radiations from the Earth crust (background radiation). Such a  ``noise" can be eliminated by employing two (or more) detectors, one above the other, that work in ``coincidence".

\item[{5.}] {\bf Scintillators.} A cosmic rays detector (telescope) is formed by different layers of scintillators. These are particular materials where a luminescence process takes place when a ionizing particle crosses it: the material absorbs the energy of the charged particle, which is then re-emitted as light (in the form of a small spark). Plastic scintillators are usually preferred, both for their ease of moulding in the desired form and for the rapidity in releasing a sufficiently strong light signal. 

\item[{6.}] {\bf Conveying the light.} The feeble light produced by the scintillator when crossed by a muon is transferred to a photomultiplier by means of an optical fiber. This is a flexible and transparent fiber made of silica or plastic, with a diameter comparable to that of a human hair, which is able to transport the light from one end to the other. The light, indeed, repeatedly bounces upon the internal walls of the optical cable, never getting it out, thanks to the phenomenon of total internal reflection. 

\item[{7.}] {\bf Producing electrons.} The light reaching the photomultiplier induces in its active material a photoelectric effect: the energy from the light emitted by the scintillator is thus used to produce electrons.  Usually, solid state photodetectors are employed for converting the light signal into an electric current. 

\item[{8.}] {\bf Avalanche multiplication.} The weak electric current produced is unable to be practically employed in order to signal the muon passage. A Silicon Photomultiplier (SiPM) is a semiconductor electronic device where the electrons, emitted when the scintillator light is absorbed, have sufficiently energy to induce ionizations, then producing further electrons in an avalanche process that drastically increases the number of electrons. The avalanche multiplication sensibly increases the photocurrent produced, resulting into an amplified electric signal that can be finally employed (for example, to turn on a LED) to detect the original muon passage.  
\end{itemize}

\section{Demonstrating cosmic rays with visualizable analogies}

\noindent After the preliminary work devoted to find and understand information about cosmic rays and their detection, the students started to elaborate a {\it hands on} model of a muon detector, suitable for understanding the relevance and the functioning of its main different parts. The strategic decision operated by the students, with the help of their Physics teacher, was to not produce a virtual model of the detector (easily traceable on the web), but rather to build a large size artifact, mainly with wood.

\begin{figure}[t]
\begin{center}
\begin{tabular}{cccc}
\includegraphics[width=3cm]{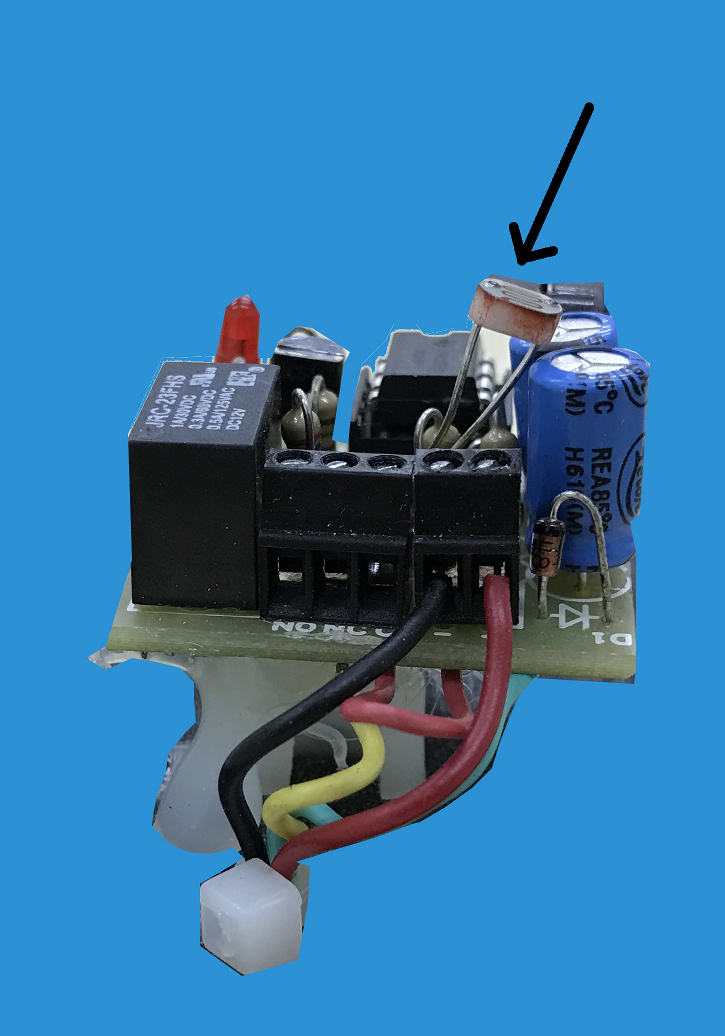} & \includegraphics[width=5.2cm]{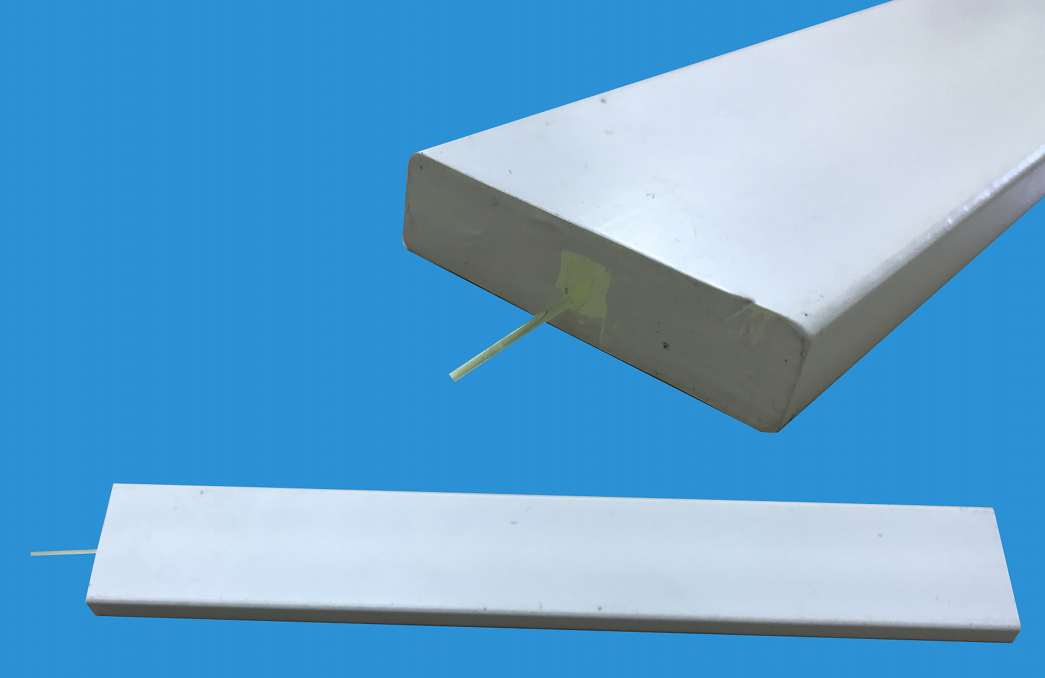} & \includegraphics[width=3cm]{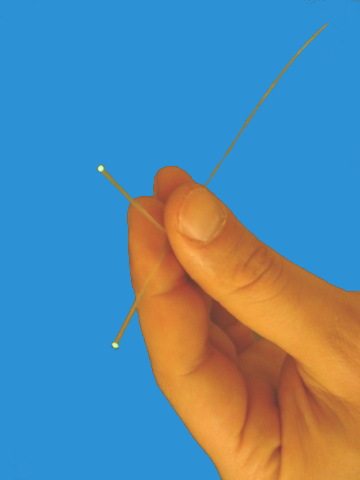} & \includegraphics[width=3cm, height=3cm]{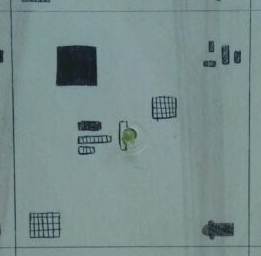} 
\end{tabular}
\caption{Some details of the demonstrator. From left to right: 1) the small integrated circuit with a semiconductor photodiode (pointed out by the arrow) employed in the model to detect the light from the incoming laser beam; 2) a plastic scintillator bar, with the optical fiber inside it; 3) an optical fiber with wavelength shifter (note, online, the greenish - rather than white - light at the ends) transmitting the light signal from the scintillator bar to the photomultiplier; 4) a front-end electronics units, with a real LED emerging from its center.}
\label{fig32}
\end{center}
\end{figure}

In the project realized, the incoming cosmic rays are visualized in terms of beams of laser light, which impinge on the detector model. This is formed by a wooden parallelepiped, upon which the typical elements of the detection front-end electronics are sketched (see Fig. \ref{fig31}). The crossed, orthogonal  superposition  (on 6 levels) of scintillator bars is particularly highlighted on the right side of the model, as well as on the top of it (see the white bars alternated with black ones in Fig.  \ref{fig31}). Just as in the real telescope, two matrices of real LEDs emerge from the right and front sides of the model (at the center of the front-end electronics units), such LED having the function of pointing out the activation of the corresponding scintillator bars when muons cross (and interact) with the detector, thus highlighting their spatial trajectory. The transmission of the light signal from the scintillators to the photomultipliers, then connected to the external LEDs, is described through the use of an optical fiber with a wavelength shifter, directly provided by the Naples' Unit of the I.N.F.N. to the students, along with a sample of a scintillator bar (see Fig. \ref{fig32}). 

\begin{figure}[t]
\begin{center}
\begin{tabular}{cc}
\includegraphics[width=7cm, height=5cm]{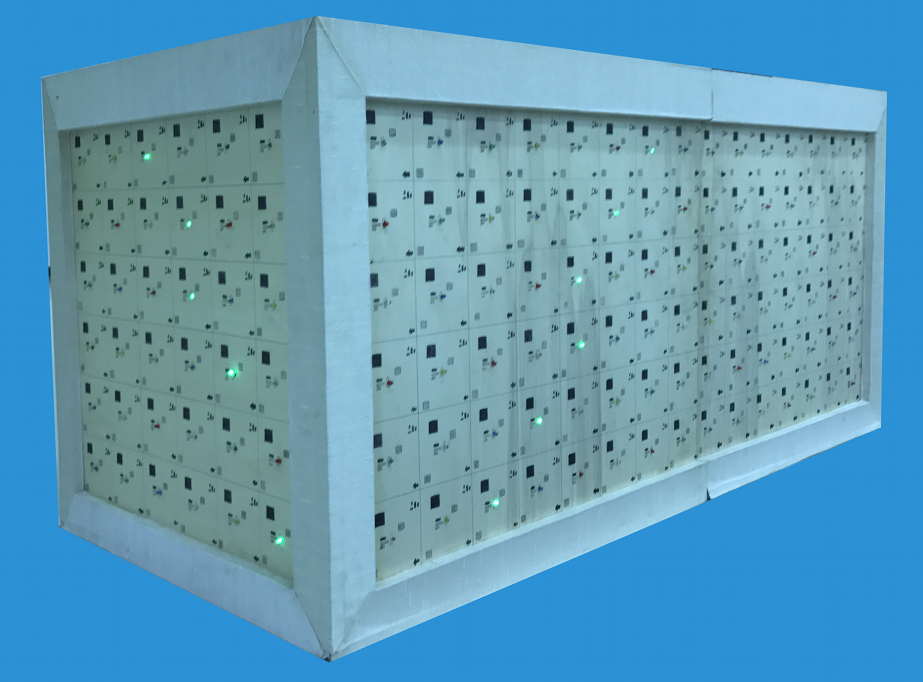} & \includegraphics[width=8cm, height=5cm]{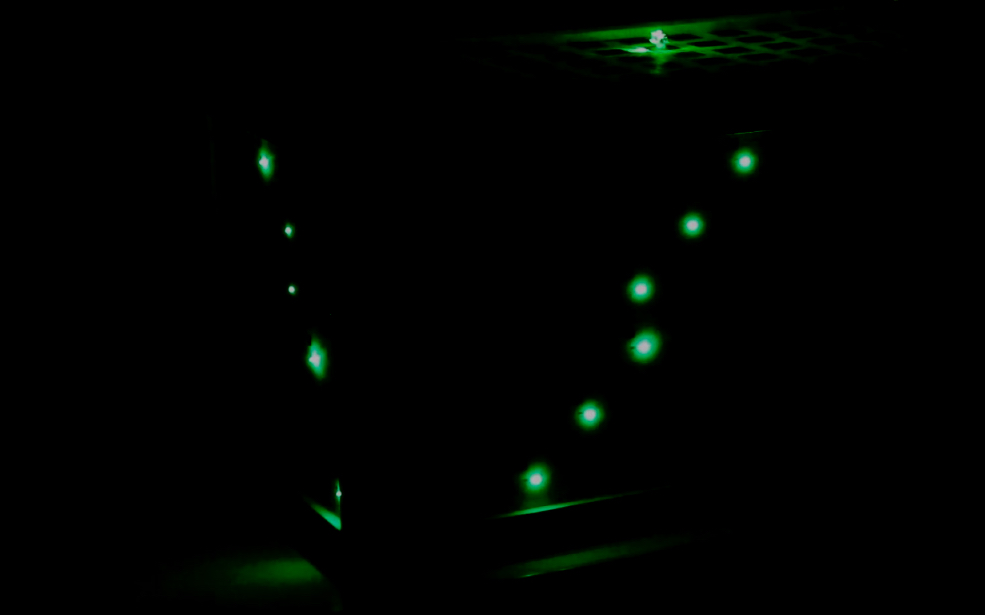}
\end{tabular}
\caption{Tracking the laser beam impinging on the demonstrator with LEDs on two sides of the model.}
\label{fig33}
\end{center}
\end{figure}

The detection of the laser beam impinging on the model -- representing an incident muon in the demonstrator -- takes place by means of a semiconductor photodiode (Fig. \ref{fig32}) placed on the top side of the model, as shown in Fig. \ref{fig31}. The laser light reaches the active part of the photodiode, thus triggering an avalanche process of electronic ionization producing an appreciable electric current. By means of an appropriate connection electronics (that can be directly seen in the interior of the model) made of electric circuits with resistors and batteries, this current is able to turn on the LEDs on the external sides of the model, thus highlighting the track of the laser beam. The three-dimensionality of the track detection is grasped through the (simultaneous) turning on of LEDs on two contiguous (orthogonal) sides of the model (Fig. \ref{fig33}).

In such a way, the general outline of the model follows strictly that of the cosmic rays telescope at the Toledo Metro Station in Naples, although the present model extends horizontally rather than vertically: as we will see below, the aim is to ``see" what is ``inside" the detector, allowing the people to understand the functioning of the different parts of it.

\section{Avalanche process in a photomultiplier}

\begin{figure}[t]
\begin{center}
\begin{tabular}{ccc}
\includegraphics[width=3.6cm, height=4cm]{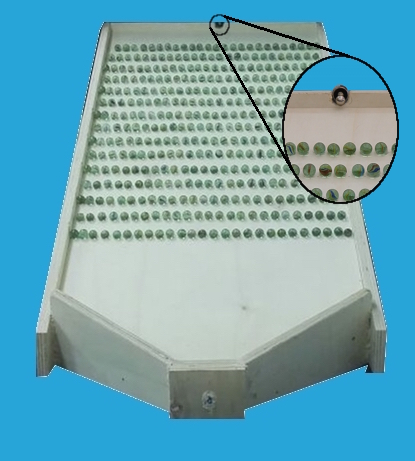} & \includegraphics[width=7.5cm, height=4cm]{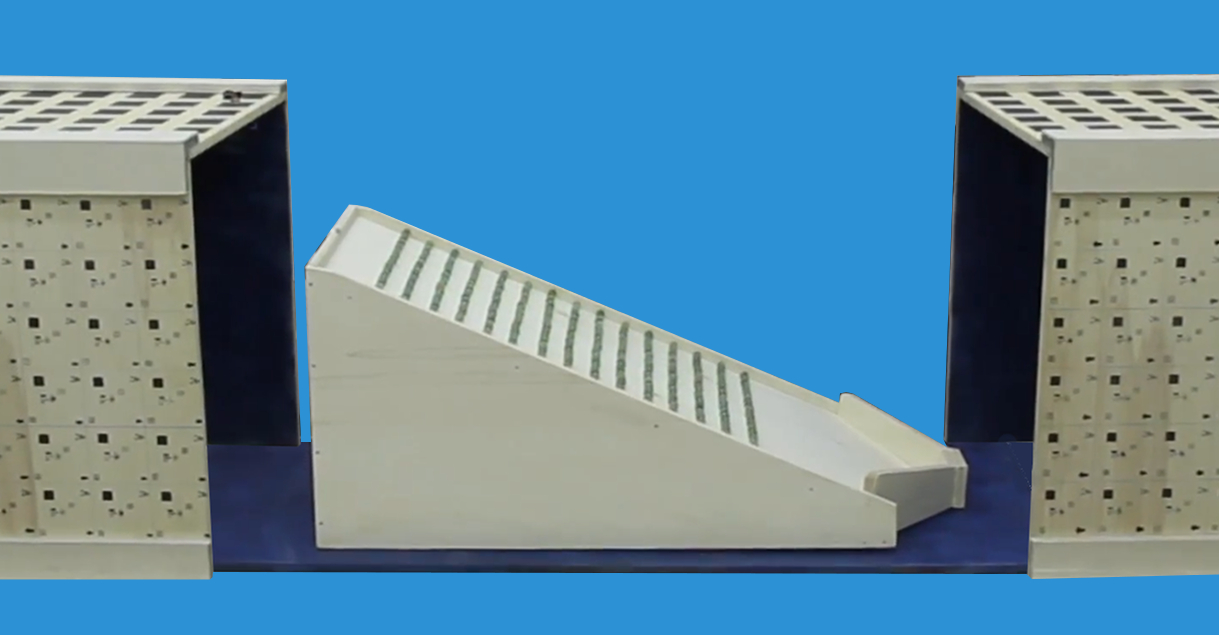} & \includegraphics[width=3.6cm, height=4cm]{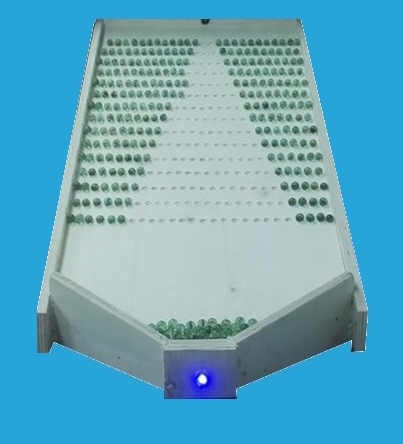}
\end{tabular}
\caption{The model describing the avalanche process of electronic ionization taking place in the photomultiplier of a cosmic rays detector. One single glass marble descending along the inclined plane is not able to turn on the LED at the bottom, while the ball released by the electromagnet on the top (see the inset) can produce an avalanche that does turn on the LED.}
\label{fig41}
\end{center}
\end{figure}

\noindent In addition to muon tracking, the present model intended to demonstrate even the avalanche process of electronic ionization that, in a photomultiplier of a cosmic rays detector (as well as in the laser photodiode in the model), leads to a suitable amplification of the single electron signal generated from the luminescent medium of the scintillator (or the semiconductor photodetector), in order to have an electric signal with sufficient energy that is able to turning on LEDs or other devices. Such avalanche process is pointed out by a second wooden model ``hidden" inside the parallelepiped model described above, having the form of an inclined plane (Fig. \ref{fig41}). Glass marbles are placed upon this inclined plane, housed inside small grooves drilled in the wood; they are able to move freely when solicited, exactly as the electrons in a SiPM. On the top of the inclined plane, a metal ball is located, kept into equilibrium by an electromagnet, which can be disabled at any moment, thus releasing the ball.

Glass marbles evidently represent the electrons that are weakly bounded within the photomultiplier (or the semiconductor), while the metal ball represents the photo-electron emitted during the photoelectric effect induced by the scintillator light (or by the laser). The metal ball released by the electromagnet (when switched off) induces an avalanche effect that kicks off the glass marbles on the inclined plane: the electric field favoring the avalanche process in the photomultiplier or the photodiode is here represented by the gravitational field of the Earth. The marbles falling along the inclined plane are then gathered on the bottom of it, where they turn on a small switch that powers a LED (see Fig. \ref{fig41}), thus pointing out -- in the present model -- how a small cause can be amplified till to produce an observable effect.

\section{Conclusions}

\noindent We have described as the detection of cosmic rays can be demonstrated to the general public, even without special physics knowledge, with the help of a detailed artifact showing all the main features of a cosmic rays telescope. A pictured poster accompanying the demonstrator revealed to be effective in assisting cosmic rays demonstration and its detection, although what most impressed the public (especially the youngest) was certainly the lighting of the LEDs showing the laser beam trajectory and, even more, the avalanche falling of glass marbles turning on a LED. Probably due to the accuracy of the two wooden demonstrators, these impressed even the learned people (including scientists and, in particular, physicists), thus contributing to break down the dividing wall between laymen and researchers. 

A short movie is available \cite{movie} where the popularization work of the six students is showed without employing heavy tones. It can certainly be used in schools and public shows, but its usefulness is likely only in inspiring other demonstrations along similar design lines. The active part of the demonstrator is the demonstrator itself, with its full hands on potential.

\section*{Acknowledgments}

\noindent The present work would never have seen the light without the fundamental contribution of the six students Daniele Aulitto, Vincenzo Jr Di Rosa, Francesco Granata, Matteo Olimpo, Francesco Panico and Pasquale Turco. The kind assistance of Dr. Paolo Mastroserio of the outreach team of the Naples' Unit of I.N.F.N. is also gratefully acknowledged, as well as that of the organizers of the competition ``A scuola di astroparticelle" (C. Aramo and M. Ambrosio) and of the Science Festival  ``Futuro Remoto".


\end{document}